\documentclass[%
reprint,
superscriptaddress,
 amsmath,amssymb,
 aps,
prb,
]{revtex4-2}

\usepackage {graphicx,epsfig,graphics,color}
\usepackage{dcolumn}
\usepackage{bm}
\usepackage{hyperref}
\usepackage{sidecap}
\usepackage{float}

\begin{document}
\title{Melting of unidirectional charge density waves across twin domain boundaries in GdTe$_{3}$}

\author{Sanghun Lee}
\altaffiliation{These authors contributed equally to this work}
\affiliation{Department of Physics, Yonsei University, Seoul 03722, Republic of Korea}

\author{Eunseo Kim}
\altaffiliation{These authors contributed equally to this work}
\affiliation{Department of Physics, Yonsei University, Seoul 03722, Republic of Korea}

\author{Junho Bang}
\affiliation{Department of Physics, Yonsei University, Seoul 03722, Republic of Korea}

\author{Jongho Park}
\affiliation{Center for Correlated Electron Systems, Institute for Basic Science, Seoul 08826, Republic of Korea}
\affiliation{Department of Physics and Astronomy, Seoul National University, Seoul 08826, Republic of Korea}

\author{Changyoung Kim}
\affiliation{Center for Correlated Electron Systems, Institute for Basic Science, Seoul 08826, Republic of Korea}
\affiliation{Department of Physics and Astronomy, Seoul National University, Seoul 08826, Republic of Korea}

\author{Dirk Wulferding} \email{dirwulfe@snu.ac.kr}
\affiliation{Center for Correlated Electron Systems, Institute for Basic Science, Seoul 08826, Republic of Korea}
\affiliation{Department of Physics and Astronomy, Seoul National University, Seoul 08826, Republic of Korea}

\author{Doohee Cho} \email{dooheecho@yonsei.ac.kr}
\affiliation{Department of Physics, Yonsei University, Seoul 03722, Republic of Korea}


\begin{abstract}

Solids undergoing a transition from order to disorder experience the proliferation of topological defects. The melting process generates transient quantum states. However, their dynamical nature with femtosecond lifetime hinders exploration with atomic precision. Here, we suggest an alternative approach to the dynamical melting process by focusing on the interface created by competing degenerate quantum states. We use a scanning tunneling microscope (STM) to visualize the unidirectional charge density wave (CDW) and its spatial progression ("static melting") across a twin domain boundary (TDB) in the layered material GdTe$_{\rm 3}$. Combining STM with a spatial lock-in technique, we reveal that the order parameter amplitude attenuates with the formation of dislocations and thus two different unidirectional CDWs coexist near the TDB, reducing the CDW anisotropy. Notably, we discover a correlation between this anisotropy and the CDW gap. Our study provides valuable insight into the behavior of topological defects and transient quantum states.



\end{abstract}

\maketitle

Topological defects are defined as singularities that appear in the order parameter space\cite{mermin1979topological}. They have been of importance in condensed matter physics due to their key role in understanding the emergence of exotic quantum phases in materials, such as those involving superconductivity\cite{agterberg2008dislocations,du2020imaging,chen2022identification,liu2023pair}, ferroicity\cite{chae2012direct,artyukhin2014landau,liu2017anomalous}, charge density wave (CDW)\cite{le2005charge,pasztor2019holographic,lim2020atomic,aishwarya2023visualizing,lafleur2023inhomogeneous}, and so on. In general, they appear when a system undergoes phase transitions that change its charge(or spin) order or symmetry\cite{saito2007topological,du2023kibble}. Their proliferation often gives rise to phenomena different from those inherent in the pristine lattice\cite{mishra2020topological}. These characteristics make topological defects highly relevant to non-equilibrium states driven by external control parameters\cite{vogelgesang2018phase,cheng2022ultrafast,wandel2022enhanced}. 

The family of \textit{R}Te$_{\rm 3}$ (\textit{R} = rare earth elements) is notable for its potential to serve as an ideal platform for exploring the mechanism of phase transitions and proliferation of topological defects, as it exhibits various broken symmetry ground states including CDW\cite{ru2008effect}, antiferromagnetic spin ordering\cite{ru2008magnetic}, and superconductivity\cite{yumigeta2021advances}. In particular, some of them are known for their unidirectional CDW, while hydrostatic pressure can generate a bidirectional CDW with a superconducting ground state\cite{hamlin2009pressure,zocco2015pressure}. Previous studies on \textit{R}Te$_{\rm 3}$ have reported that light pulses can also induce melting of the equilibrium CDW state and the temporal evolution of an out-of-equilibrium CDW\cite{schmitt2008transient, zong2019evidence,kogar2020light,zhou2021nonequilibrium}. These transient states host the bidirectional CDW demonstrated by the pump-and-probe diffraction\cite{zong2019evidence,kogar2020light,zhou2021nonequilibrium}. It has been proposed that the phase transitions occurring in such non-equilibrium regions are closely related to the proliferation of topological defects\cite{kosterlitz1973ordering}. However, since it is challenging to acquire microscopic details of the ultra-fast transient states, information about the spatial distribution of the topological defects, its relation with the order parameter amplitude, and the electronic structure of the excited states has been considerably limited so far. 

Here, we present the spatial variations of CDW across twin domain boundaries (TDB), where two orthogonally aligned unidirectional CDW domains compete. We utilized low-temperature scanning tunneling microscopy and spectroscopy (STM/S) to visualize the charge modulations and electronic structure across the domain boundary. Our analysis employs a spatial lock-in technique to demonstrate the static CDW melting with a spatial gradient of its order parameter across the TDB and the proliferation of topological defects. We also find that two unidirectional CDWs coexist at the TDB, and that the normalized difference in their amplitude anti-correlates with the density of states (DOS) close to the Fermi level ($E_{\rm F}$). These findings offer a microscopic understanding of the formation of topological defects and their accompanying melting process, while also suggesting a connection to the superconducting phase of this system. 

\begin{figure*}
\includegraphics[scale=1]{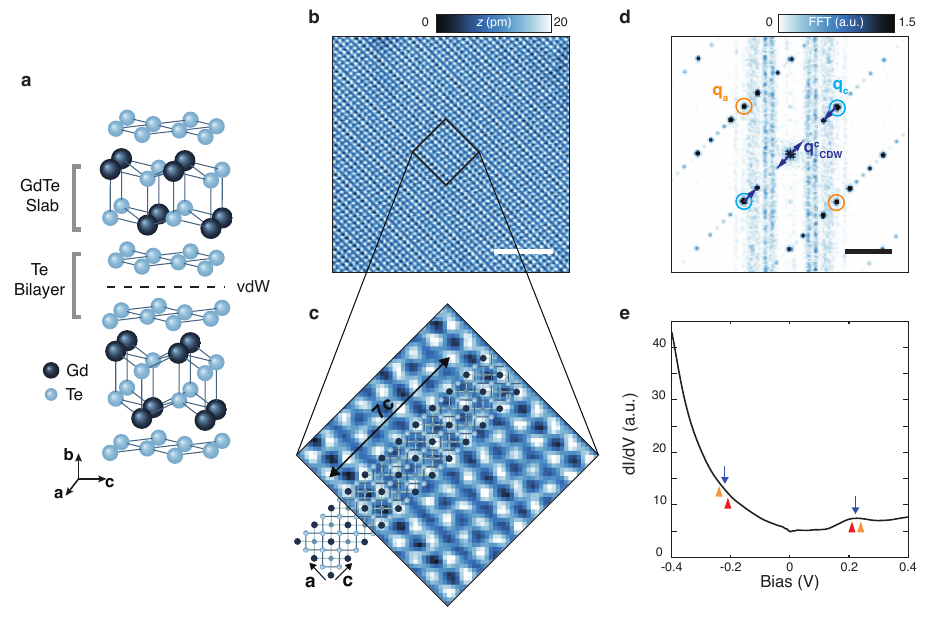}
  \caption{Unidirectional CDW in GdTe$_{\rm 3}$. (a) Crystal structure of GdTe$_{\rm 3}$. The semiconducting GdTe slab is encapsulated between conducting Te layers. The dashed line between Te layers indicates the cleavage plane. (b) An STM topographic image of the cleaved surface of GdTe$_{\rm 3}$ acquired with $V_{\rm set} = -100$ mV and $I_{\rm set} = 100$ pA.  $Scale$ $bar$, 5 nm. (c) Zoomed-in view of the area marked by the black square in (b). The Gd lattice is resolved with a unidirectional CDW propagating along the $c-$axis. (d) FFT of the STM topographic image shown in (b). Orange and light-blue circles indicate the Bragg peaks of Gd lattice along $a-$ and $c-$axes. Dark blue arrows are associated with the wavevectors of the unidirectional CDW along the $c-$axis. $Scale$ $bar$, 1/\AA. (e) The $dI/dV$ tunneling spectrum of GdTe$_{\rm 3}$ taken with $V_{\rm set} = -400$ mV, $V_{\rm mod} = 10$ mV, and $I_{\rm set} = 1$ nA. Two dark blue arrows mark the band edges of the partially opened CDW gap. One in the empty state is determined by the spectral peak, and the other in the filled state by an assumption of particle-hole symmetry. To make a comparison, the gaps identified by STM \cite{lei2020high, xu2021molecular} and optical spectroscopy \cite{hu2014coexistence} are marked with red and orange triangles, respectively. All color bars are linear scales.}
  \label{Fig.1}
\end{figure*}

GdTe$_{\rm 3}$ is a layered material comprising two conducting square Te nets separated by an insulating GdTe slab, as illustrated in Figure \ref{Fig.1}(a). The Te layers are weakly coupled by van der Waals interactions, suggesting that a Te layer terminates each cleavage plane. This metallic surface layer is associated with the unidirectional CDW observed in previous STM studies. An STM topographic image of the cleaved surface, shown in Figure \ref{Fig.1}(b), reveals a square lattice whose unit cell is consistent with the Gd lattice. This can be rationalized by its protrusion closer to the surface and its higher charge density  compared to Te atoms in the insulating GdTe layer, resulting in a higher STM current \cite{tomic2009scanning}. The image also exhibits regular stripe patterns with wavefronts propagating along the $c-$axis. This unidirectional charge modulation has been known as an incommensurate CDW phase with a wavelength of about $7/2c$, where $c$ is the lattice constant along the direction perpendicular to the wavefronts. Superimposing the CDW superstructure with the Gd lattice results in unidirectional patterns with a periodicity of about $7c$, as indicated by the black arrow in Figure \ref{Fig.1}(c). 

The Bragg peaks for the Gd lattice and the unidirectional CDW are clearly visible in the fast Fourier transform (FFT) of the STM topographic image. The Bragg peaks for the pristine lattice are highlighted by orange and light-blue circles in Figure \ref{Fig.1}(d). Additionally, the satellite peaks marked by blue arrows correspond to the unidirectional CDW. As shown in Figure \ref{Fig.1}(a), the bulk GdTe$_{\rm 3}$ has a glide symmetry along the $c-$axis rather than the $a-$axis, with the mirror plane between the Te layers. Due to the anisotropy in the crystal symmetry, the lattice constant $c$ is slightly larger than $a$, and thus the unidirectional CDW forms preferably along the $c-$axis\cite{yumigeta2021advances}. Therefore, the orientation of the CDW may serve as a reference for determining the direction of the lattice vectors. 
By fitting the Bragg peaks in the FFT, we determine that the lattice constants is $a(c)\sim4.3$ \AA, and the incommensurate CDW wavevector $\textbf{q}_{\rm CDW}\sim 2/7$\textbf{q}$_{\rm c}$. These results are consistent with previous studies\cite{chen2019raman}.

The differential conductance ($dI/dV$) spectrum (Figure \ref{Fig.1}(e)) allows an estimate of the CDW gap of about $2\Delta \approx 420$ meV\cite{lei2020high,xu2021molecular,hu2014coexistence}, which is determined by assigning the band edges marked with dark blue arrows. In addition, the derivative of the spectrum highlights a subtle change around the assigned edge of the valence band (See Figure 5). In 2D metallic systems, a CDW gap opens when the states are connected with the nesting vector corresponding to the CDW wavevector (\textbf{q}$_{\rm CDW}$). Therefore, a CDW state in 2D metallic systems is likely to have a partially gapped Fermi surface, resulting in the finite DOS at $E_{\rm F}$. The Fermi surface of GdTe$_{\rm 3}$ exhibits a diamond-like structure, which is the result of the combination of Fermi surfaces originating from two perpendicular 1D metallic chains. Those are attributed to the inter-site hopping of $p_{\rm x}$ and $p_{\rm y}$ orbitals and the weak inter-orbital interaction between them in the Te square net \cite{laverock2005fermi, yao2006theory}. This Fermi surface topology satisfies the unidirectional nesting condition with band-folding driven by the subsurface Gd-lattice. In accordance with our data, recent studies using angle-resolved photoemission spectroscopy (ARPES) have revealed an anisotropic gap opening and the remnant Fermi surface at low temperatures\cite{lei2020high, liu2020electronic}. In addition, charge modulations between the conduction and the valence band edge exhibit an out-of-phase relation, which has been commonly observed in Peierls-type CDW systems\cite{dai2014microscopic,spera2020insight,pasztor2021multiband,jang2022direct} (See Figure 6).\bigskip

\begin{figure*}
\includegraphics[scale=1]{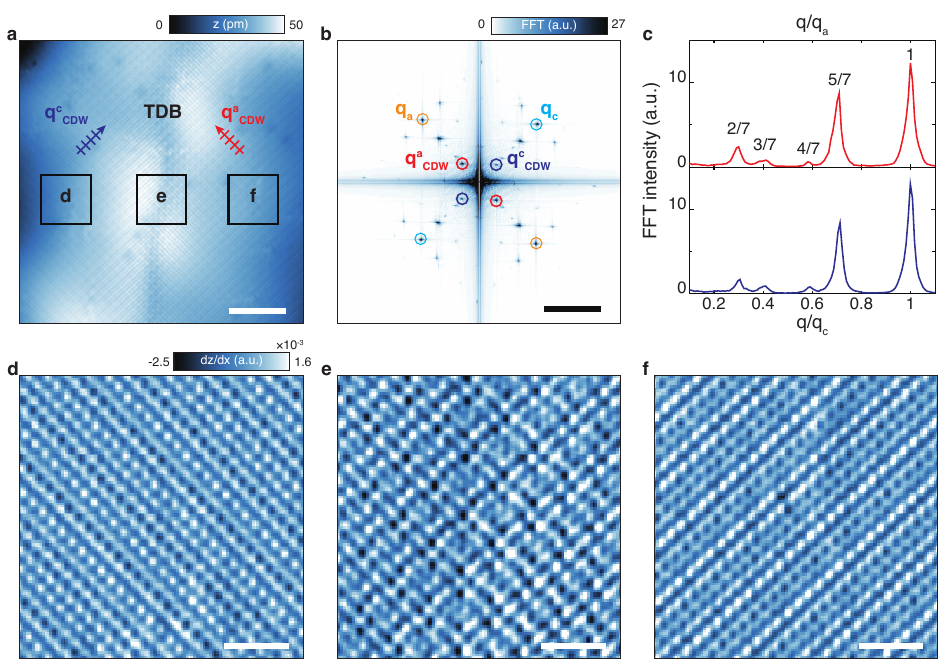}
\caption{A TDB with perpendicularly oriented unidirectional CDWs. (a) An STM topographic image of two different unidirectional CDW domains and a TDB acquired with $V_{\rm set} = 400$ mV and $I_{\rm set} = 300$ pA. $Scale$ $bar$, 10 nm. (b) FFT of the STM topographic image shown in (a). Orange and light-blue circles highlight the Bragg peaks of the Gd lattice. Red and blue circles are associated with the Bragg peaks of the unidirectional CDW propagating along the $a-$ and $c-$axes, respectively. $Scale$ $bar$, 1/\AA. (c) Line profiles of the FFT intensity in (b) along $a-$ (red line) and $c-$axes (blue line). The wavevectors are normalized with the reciprocal lattice vectors for the same direction. (d-f) Zoomed-in view of the area marked by the white squares in (a). $Scale$ $bar$, 2 nm. All maps in (d-f) were spatially differentiated. All color bars are linear scale.}\label{Fig.2}  
\end{figure*}

\begin{figure*}
\includegraphics[scale=1]{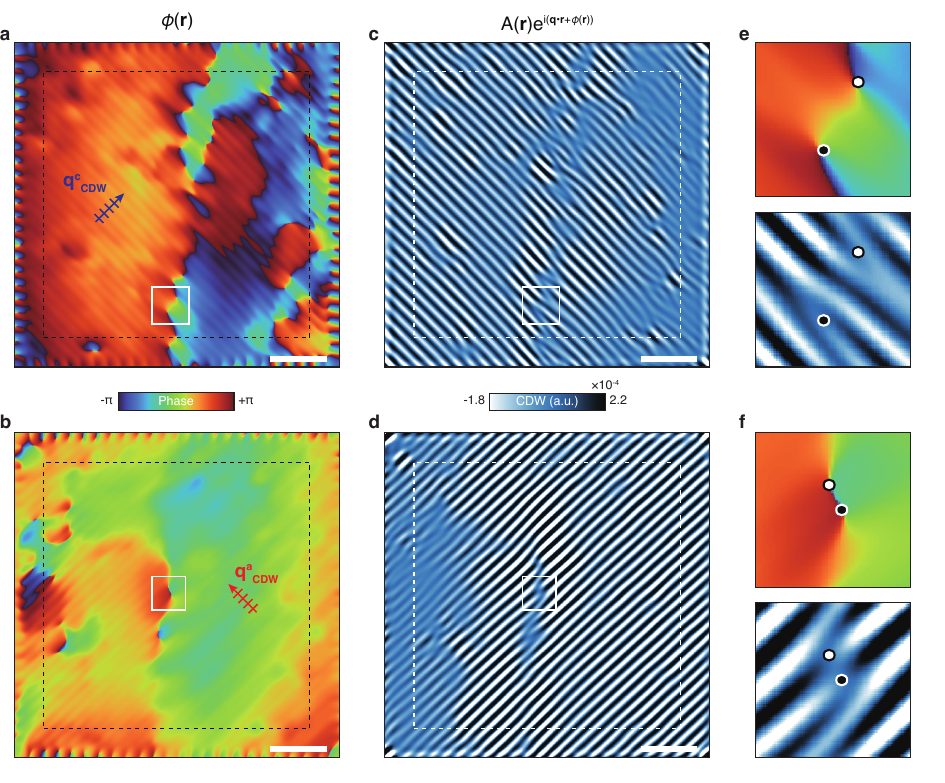}
\caption{Visualizing amplitude and phase of the unidirectional CDW. (a) and (b) Phase maps for the unidirectional CDWs along the $c-$ and $a-$axes, respectively, obtained by the spatial lock-in technique. (c) and (d) The CDWs reconstructed by combining the amplitude and phase acquired with reference signals corresponding to the unidirectional CDW along $c-$ and $a-$ axes. (e) and (f) The zoomed-in images of topological defect pairs in the area marked with white squares in (a(c)) and (b(d)), respectively. The singular points where phase winding occurs in $\pm2\pi$, highlighted with black (clockwise) and white (counterclockwise) dots, are characterized by the CDW dislocations. All maps in (a-d) were acquired within the same field of view as the STM topographic image in Figure \ref{Fig.2}(a). $Scale$ $bar$, 10 nm. All color bars are linear scale.}\label{Fig.3}  
\end{figure*}

X-ray diffraction\cite{banerjee2013charge}, transport\cite{ru2008effect}, and STM\cite{lei2020high,xu2021molecular} studies have reported that GdTe$_{\rm 3}$ has a unidirectional CDW below $T_{\rm c}\sim 340$ K. This anisotropy has been attributed to the broken $C_4$ rotation symmetry (|$\mathbf{q}_{\rm a}$| > |$\mathbf{q}_{\rm c}$|) inherent in the crystal structure. We confirm this symmetry breaking by additional polarization-resolved Raman spectroscopy measurements\cite{wang2022axial}, which can directly probe the orientation of unidirectional CDW domains on a larger micrometer-scale (see Figure 7). However, external control parameters, such as strain\cite{straquadine2022evidence} or laser pulses\cite{kogar2020light}, may enable the coexistence of two perpendicularly oriented unidirectional CDW states domain by domain. These degenerate CDWs share the same crystallographic structure but propagate along orthogonal directions. The CDW domains are separated by the interface, which we have previously referred to as twin domain boundaries (TDB). In the following, we outline our observation and characterization of such TDBs.

Our STM topographic image (Figure \ref{Fig.2}(a)) illustrates the presence of the TDB, where two CDW domains with perpendicular orientations intersect. The arrows with ticks in Figure \ref{Fig.2}(a) indicate the direction of propagation and wavefronts of the CDW for each domain, respectively. Notably, the TDB in the center line of the STM topographic image exhibits a bidirectional CDW resulting from the superposition of two unidirectional CDWs. The differentiated STM images (Figure \ref{Fig.2}(d)-(f)) provide a clearer representation of two distinct CDWs and their coexistence at the TDB. These unidirectional CDWs are not sharply separated but spread across the TDB with amplitude attenuation.

In Figure \ref{Fig.2}(b), the FFT of our STM topographic image shows two sets of distinct CDW peaks near the $\Gamma$ point, each representing unidirectional CDWs along the $a-$axis (red circles) and $c-$axis (blue circles), respectively. Further measurements demonstrate that the CDW along the $c-$axis is dominant throughout the entire surface (see Figure 8), while the CDW along the $a-$axis is locally formed. It has been well established that the direction of the CDW is determined by the anisotropy of the pristine lattice\cite{yumigeta2021advances}. However, as shown in Figure \ref{Fig.2}(b), the Bragg peaks of the Gd lattice, highlighted by orange and light-blue circles, are sharply defined without peak splitting since the structural anisotropy is beyond our experimental resolution. 

To compare two perpendicular CDWs, we acquired line profiles of the FFT intensity along $a-$ and $c-$axes in Figure \ref{Fig.2}(c). The wavevectors are normalized with the reciprocal lattice vectors of GdTe$_{\rm 3}$. They commonly show the strongest peak and four additional peaks at 1, 2/7, 3/7, 4/7, and 5/7, corresponding to $\mathbf{q}_{\rm a(c)}$,  $\mathbf{q}_{\rm CDW}$, 1-2$\mathbf{q}_{\rm CDW}$, 2$\mathbf{q}_{\rm CDW}$, and 1-\textbf{q}$_{\rm CDW}$, respectively. The normalized two different CDW wavevectors are identical within the resolution of our data. 

When examining the spatial variation across the TDB, we can observe a gradual diminishing of the CDW in each domain as it crosses into the other domain. The process, which is termed "static melting" in this paper, can be accompanied by a reduction of the CDW amplitude and a twist of its phase. To acquire the spatial variations of the amplitude($A_{\rm CDW}(\mathbf{r})$) and the phase(${\phi_{\rm CDW}(\mathbf{r})}$) of the unidirectional CDW, we employed a spatial lock-in technique (see details in Figures 9 and 10). Figure \ref{Fig.3}(a) and (b) display changes in the phase of the individual unidirectional CDWs, which propagate along the $c-$ and $a-$axes, respectively. We note that the CDWs are almost coherent in their domains, while several phase windings of $\pm2\pi$, so-called topological defects, are observable across the TDB. 

Topological defects in the form of $2\pi$ phase winding are known to induce dislocations in the CDW wavefront \cite{mermin1979topological}. We can construct the individual unidirectional CDWs with $A_{\rm CDW}(\mathbf{r})$ and ${\phi_{\rm CDW}(\mathbf{r})}$ acquired by the lock-in technique. Figure \ref{Fig.3}(c) and (d) clearly show the attenuation of the CDW amplitude and the proliferation of dislocations across the TDB. These results suggest that the phase stiffness of the unidirectional CDW can be reduced when its order parameter amplitude decreases\cite{emery1995importance}. In addition, by comparing the spatial relationship between their phase and amplitude, it is evident that phase singularities occur where the CDW amplitude is strongly reduced (Figure \ref{Fig.3}(e) and (f)). The local excitation is unstable against the CDW reconstruction, either restoring the initial state or creating the $90^\circ$ rotated unidirectional CDW. In that case, the system exhibits the bidirectional CDW\cite{kogar2020light}. 

\begin{figure*}
\includegraphics[scale=1]{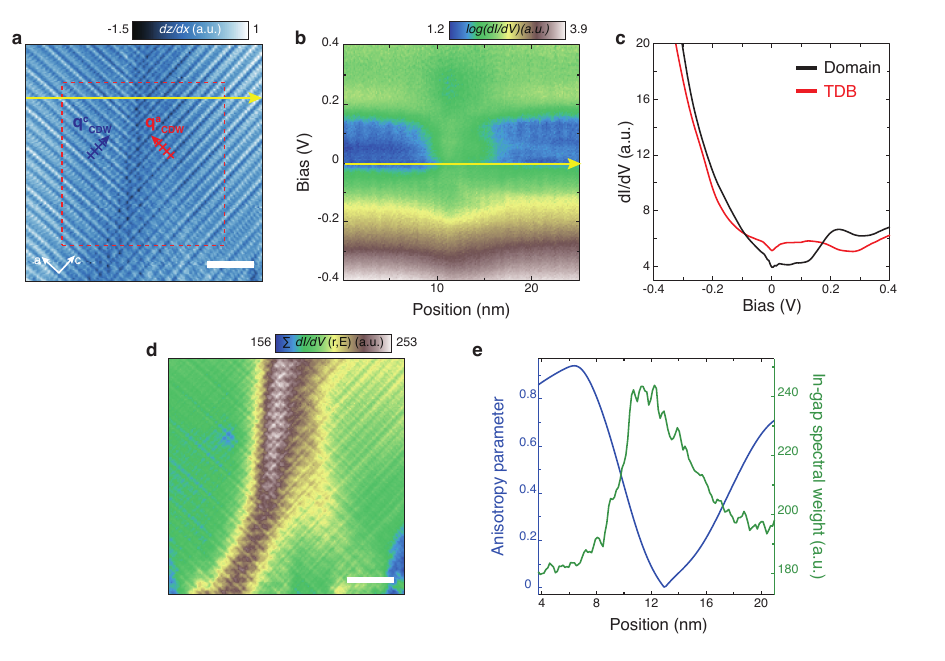}
\caption{Electronic structure of the TDB. (a) The spatial derivative of the STM topographic image of a TDB region acquired simultaneously with the two-dimensional spectroscopic mapping. $Scale$ $bar$, 5 nm. (b) The line cut of $dI/dV$ map along a yellow arrow in (a). (c) Averaged spectra acquired in the TDB (red line) and in the CDW domain (black line). Spectral weight transfer leads to the in-gap like feature close to $E_{F}$ and also suppression of the DOS around +300 meV. In addition, no gap opening is apparent at higher binding energy. (d) In-gap spectral weight map of the area marked by the red dashed square in (a). In-gap spectral weight is obtained by integrating $dI/dV$ from $-100$ to $+100$ meV. Tunneling conditions for $dI/dV$ mapping are $V_{\rm set} = -400$ mV and $I_{\rm set} = 1$ nA. $Scale$ $bar$, 5 nm. (e) Anti-correlation between the anisotropy parameter and the in-gap spectral weight along the line marked by a yellow arrow in (a). All color bars are linear scale.}\label{Fig.4}  
\end{figure*}

The gradual attenuation of each unidirectional CDW amplitude across the TDB allows us to investigate the difference between unidirectional and bidirectional CDWs shown in Figure \ref{Fig.4}(a). We performed spectroscopic mapping in the vicinity of the TDB. Figure \ref{Fig.4}(b) shows differential conductance ($dI/dV$) spectra acquired along the yellow arrow in Figure \ref{Fig.4}(a). The $dI/dV$ spectra averaged over the TDB and the domains are compared in Figure \ref{Fig.4}(c). The direct comparison demonstrates two notable distinctions: a reduced CDW gap and an enhanced DOS near $E_{\rm F}$ at the TDB. Similar observations have been reported in other \textit{R}Te$_{\rm 3}$ samples, in cases where checkerboard CDW is formed through phase transitions\cite{hu2014coexistence}, or when a bidirectional CDW is induced by laser pulses\cite{zong2019evidence}.

We further analyze the spatial evolution of DOS near $E_{\rm F}$ by calculating the in-gap spectral weight ($\Sigma dI/dV$). For a direct comparison with previous time-resolved ARPES studies\cite{zong2019evidence}, we integrate the differential conductance from $-100$ meV to $+100$ meV, which is within the CDW gap. As shown in Figure \ref{Fig.4}(d), the in-gap spectral weight exhibits a local enhancement at the TDB, while the bidirectional CDW states exist across a spatially extended region and topological defects are scattered in the proximity of the TDB. To correlate the localized spectral feature with the CDW order parameter, we determined the amplitude of the CDW, $A_{\rm a}$ and $A_{\rm c}$, as a function of position (see details in Figure 11). Subsequently, we define an anisotropy parameter $\rm{AP}$ as follows: $\rm{AP} = \left\vert\frac{A_{\rm c} - A_{\rm a}}{A_{\rm c} + A_{\rm a}}\right\vert$. The anisotropy parameter is close to 1 for the unidirectional CDW and 0 for the bidirectional (checkerboard-like) CDW. As shown in Figure \ref{Fig.4}(e), the anisotropy parameter strongly anti-correlates with the in-gap spectral weight in Figure \ref{Fig.4}(d).

Note that such bidirectional CDWs in $R$Te${_3}$ have been commonly observed in excited states driven by external stimuli with topological defects\cite{zong2019evidence, kogar2020light}. They have been reported to have higher LDOS than their ground states around $E_{F}$. However, these quantum states are neither static nor spatially homogeneous, which hinders an accurate determination of their atomic and electronic properties. In this regard, we highlight the TDB, which can be locally pinned by strain gradients and exist at the interface between two perpendicularly aligned unidirectional CDWs. Examining the TDB allows us to observe the decay behavior statically of the order parameter amplitude and phase, which is likely to exhibit a charge order similar to that of the excited state. Remarkably, we find that the electronic structure is primarily associated with the relative difference of the CDW amplitude between degenerate ground states, rather than with their phase. This is consistent with the decay behavior of the laser-induced excited states in previous time-resolved ARPES experiments~\cite{zong2019evidence}, highlighting that the restoration of the CDW gap occurs much faster than its phase coherence. Therefore, this approach is an ideal tool for investigating the electronic properties of local excitations, such as bidirectional CDW or topological defects, to better understand short-lived quantum states out of equilibrium. 

In conclusion, our STM results reveal the spatial progression of the unidirectional CDW close to a TDB in GdTe$_{\rm 3}$. Our spatial lock-in technique enables us to find the local excitations and their distribution which so far have been extensively discussed for understanding the dynamical melting process of the unidirectional CDW without microscopic experimental evidence. We note that the amplitude of the CDW order parameter decreases with winding its phases, thus constructing the unidirectional CDW wavefronts' dislocations and the bidirectional CDW. This intermediate region where two unidirectional CDWs coexist exhibits the enhanced DOS close to $E_{\rm F}$, which was observed in pump-probe photoelectron spectroscopy studies~\cite{zong2019evidence}. The higher electron density near $E_{\rm F}$, accompanied by a bidirectional CDW, might be connected to the emergence of superconductivity in the bidirectional (or checkerboard) CDW in \textit{R}Te$_{\rm 3}$ systems~\cite{ru2008effect,fang2019disorder,fang2020robust}. Lastly, our approach will be instructive to study competing phenomena between charge(or spin)-ordered phases and superconductivity in low-dimensional correlated systems. To gain a deeper understanding of this complex relationship, it is required to conduct additional STM experiments on the systems that exhibit the charge order with superconductivity below the critical temperatures. These atomically resolved experiments should focus on investigating the connection between the CDW and the superconducting order parameter on the nanoscale~\cite{liu2021discovery}.

\begin{acknowledgments}
The authors thank M. P. Allan, T. Benschop, J. -F. Ge, G. Kim, S, Kim, J. Rhim, S. Park, and Y. Son for valuable discussions. S. L., E. K., J. B., and D. C. were supported by the National Research Foundation of Korea (NRF) grant funded by the Korea government (MSIT) (No. 2020R1C1C1007895, 2017R1A5A1014862, and RS-2023-00251265) and the Yonsei University Research Fund of 2019-22-0209. J. P., C. K., and D. W. acknowledge support from the Institute for Basic Science (IBS) (Grants No. IBS-R009-G2, IBS-R009-Y3).
\end{acknowledgments}

\newpage
\section{Methods}
\noindent{\textbf{Sample growth.}} We used stoichiometric single-crystalline GdTe$_{\rm 3}$ flakes to grow crystals through the self-flux method. We mixed high-purity Gd metal (99.9\%) and Te chips (99.999\%) in a 1:30 molar ratio for the Te-rich self-flux condition. The mixed precursor of GdTe$_{\rm 3}$ was loaded into quartz tubes and sealed at $\sim$10$^{-5}$ Torr with a turbo pump to avoid oxygen contamination. We heated the mixture to a maximum temperature of 900$^\circ$C and maintained the temperature for 24 hours to achieve a homogeneous melt. Afterward, we cooled the sample down to 500$^\circ$C at a rate of $-$2$^\circ$C/hour. We decanted the melted mixture in an ampule and waited for it to solidify at room temperature. The crystal structure was determined using a high-resolution X-ray diffractometer with an airtight holder to prevent oxygen-caused degradation.

\noindent{\textbf{STM/S measurements.}} STM/S measurements were carried out using a commercial low-temperature STM (UNISOKU) at liquid helium temperature ($T=4.2$ K) in ultra-high vacuum (UHV) environment down to $1 \times 10^{-10}$ Torr. The single crystal samples of GdTe$_{\rm 3}$ were exfoliated \textit{in situ} at room temperature and loaded to the precooled STM head. Mechanically sharpened Pt-Ir (90/10) wires were used for STM tips. STM topographic images were obtained in constant current mode by applying bias voltages ($V_{\rm set}$) to the sample. The differential conductance ($dI/dV$) images were acquired using a standard lock-in technique, with an a.c. voltage (a peak-to-peak voltage amplitude $V_{\rm pp} = 10$ mV and a frequency $f_{\rm mod} = 613$ Hz added to the d.c. sample bias.

\noindent{\textbf{Raman scattering measurements.}} Raman scattering measurements shown in Figure 7 were carried out at room temperature (295 K) with a single mode laser (Oxxius, 561 nm). To avoid degradation of the surface, the sample was freshly cleaved, installed, and sealed into the sample chamber while inside a glove box. After removing the sealed sample chamber from the glove box, we evacuated it with a turbo pump, thereby avoiding any air contamination. The polarization of the incident laser light was rotated with a superachromatic $\lambda/2$ waveplate within the crystallographic $ac$-plane. For the polarization-resolved Raman maps, the polarization of the incident light and the polarization of the scattered light entering the spectrometer are at a fixed parallel relation to each other ("parallel polarization"). The inelastically Raman-scattered light was dispersed through a Princeton Instruments TriVista spectrometer and detected via a PyLoN eXcelon charge-coupled device.

\section{Supporting Information}

\begin{figure*}
\includegraphics[width=14cm]{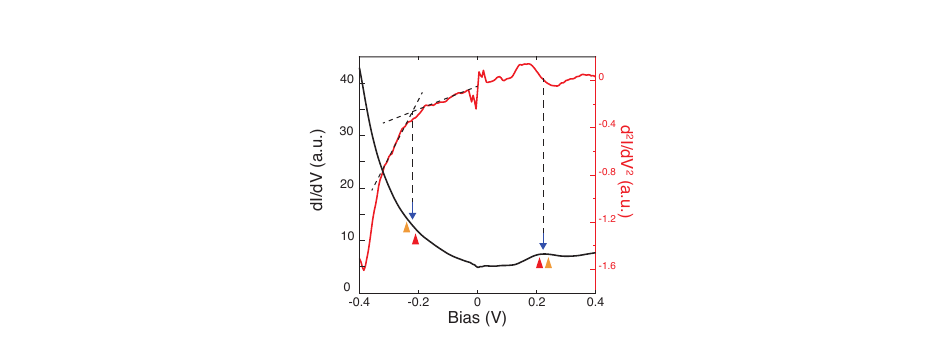}
\caption{\textbf{Assignment of the CDW gap in the tunneling spectrum.} Due to partial gap opening, the CDW gap does not show the zero conductivity in the differential tunneling spectrum. In the main text, we show the spatially averaged tunneling spectrum in Figure 1(e). We assign the small peak (marked by a dark blue arrow) in the empty state as the edge of the conduction band. Due to the lack of a distinct shoulder-like feature in the filled state, we determine the band edge at the energy where the rate of increase in the DOS changes, shown in its derivative (red curve) and highlighted by the dashed black lines.}\label{EFig.1}  
\end{figure*}

\begin{figure*}
\includegraphics[width=12cm]{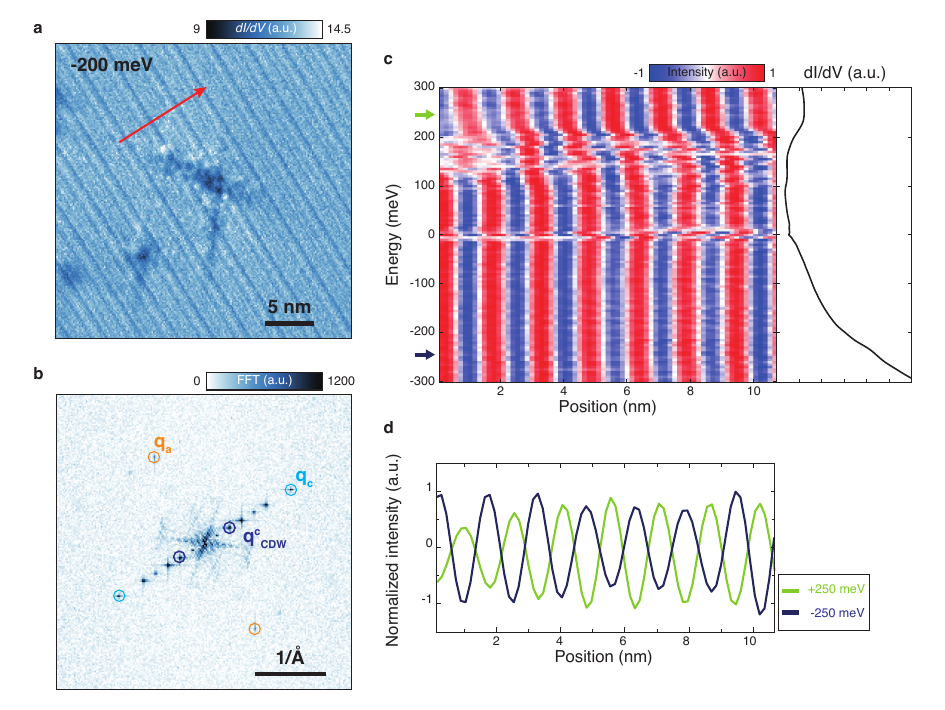}
\caption{\textbf{Phase relation of the energy dependent charge modulations in the unidirectional CDW.} \textbf{a,} A $dI/dV$ map of the unidirectional CDW at the energy of -200 meV. It shows the charge modulations and the quasiparticle interference from the defects. \textbf{b,} The FFT spectrum of the $dI/dV$ map shown in \textbf{a}. This spectrum highlights the wavevectors of the lattice and the unidirectional CDW, with colored circles indicating their positions. \textbf{c,} The energy-dependent $dI/dV$ line profiles. They are acquired along the perpendicular direction (red arrow in \textbf{a}) of the CDW wavefronts. All profiles have been normalized for clarity. \textbf{d,} Phase relation of the charge modulations in valence and conduction bands. Notably, the profiles at the edges of the valence and conduction bands exhibit an out-of-phase relation, a common feature observed in Peierls-type CDW systems. To gain a deeper understanding of energy-dependent charge modulations, theoretical calculations with the multiband structure, partially opened energy gap, and incommensurate charge modulations superimposed on atomic corrugations are required.}\label{EFig.2}  
\end{figure*}

\begin{figure*}
\includegraphics[width=14cm]{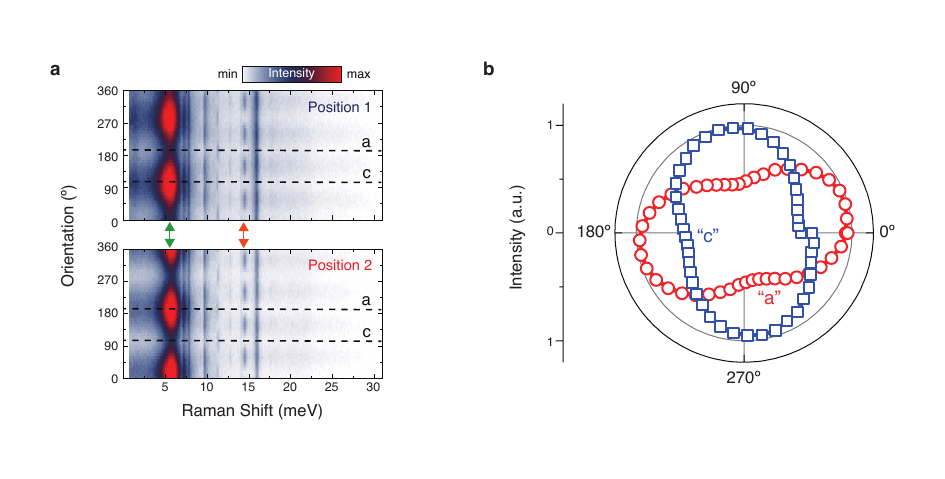}
\caption{\textbf{Unidirectional CDW and its rotation.} \textbf{a,} The prominent CDW amplitude mode (marked by a green arrow) observed with Raman spectroscopy and its distinct 2-fold symmetry in the polarization-resolved Raman maps. A four-fold symmetric phonon mode (marked by an orange arrow) used to assign crystallographic $a$ and $c$ axes, indicated by dashed black lines. Two different measurement positions, a few tens $\mu$m apart from each other, show similar behavior, although the amplitude mode is offset by a 90$^{\circ}$ phase between these two positions "1" and "2". \textbf{b,} Angle-dependent integrated Raman scattering intensity of the CDW amplitude mode at position "1" (blue squares) and position "2" (red circles). Note that the diameter of the laser spot (i.e., the surface area from which the signal was collected) was about 2 $\mu$m. The observed two-fold symmetry of the CDW amplitude mode can be direct evidence that larger, $\mu$m-sized domains exist in GdTe$_3$.}\label{EFig.3}  
\end{figure*}

\begin{figure*}
\includegraphics[width=14cm]{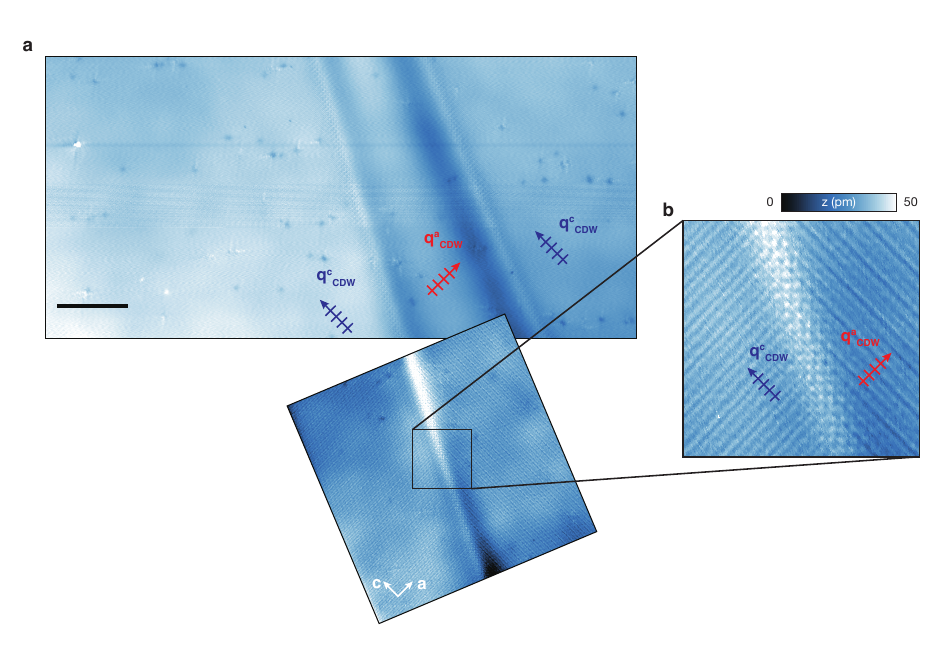}
\caption{\textbf{Coexistence of two different unidirectional CDWs.} \textbf{a,} Large-scale STM topographic image including twin domains that have perpendicularly oriented unidirectional CDWs. One domain with the CDW propagating along the $c-$axis is larger than the other. \textbf{b,} A zoomed-in STM topographic image of the TDB.}\label{EFig.4}
\end{figure*}

\begin{figure*}
\includegraphics[width=12cm]{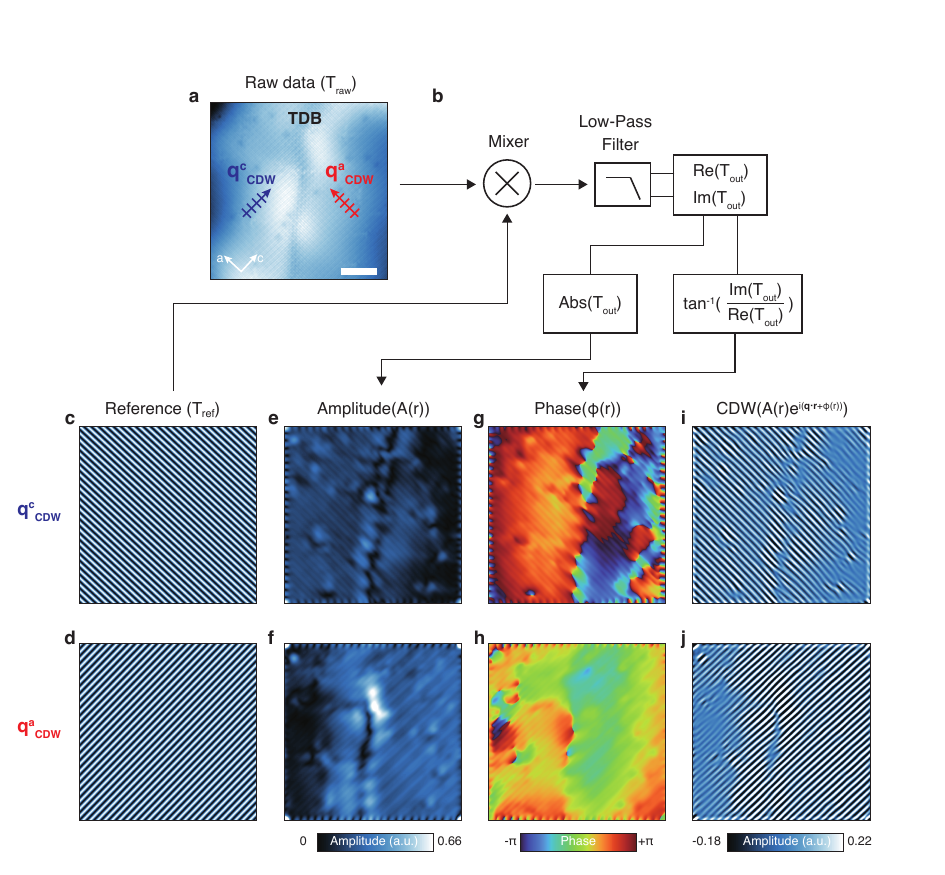}
\caption{\textbf{CDW phase analysis with spatial lock-in technique.} \textbf{a,} An STM image of the TDB shown in Figure 2(a): $T_{\rm raw} = \sum_{\rm j}\frac{A_{\rm j}(\mathbf{r})}{2}({\rm e}^{\mathrm{i}[\mathbf{q}_{\rm j}\cdot\mathbf{r}+\phi_{\rm j}(\mathbf{r})]}+\mathrm{e}^{-\mathrm{i}[\mathbf{q}_{\rm j}\cdot\mathbf{r}+\phi_{\rm j}(\mathbf{r})]})$, where $\phi_{\rm j}(\mathbf{r})$ is the position-dependent phase. \textbf{b,} Block diagram of the spatial lock-in algorithm composed of a signal mixer and a low-pass filter. \textbf{c} and \textbf{d,} The reference signals corresponding to the unidirectional CDWs: $T_{\rm ref}={\rm e}^{-\mathrm{i}\mathbf{q}_{\rm CDW}\cdot\mathbf{r}}$. As shown in \textbf{b}, $T_{\rm raw}$ is multiplied by $T_{\rm ref}$ with the mixer. The subsequent low-pass filtering can extract the DC component from the mixing signal. In the case of equal wavevector $\mathbf{q}_{\rm j}=\mathbf{q}_{\rm CDW}$, the lock-in output given by $\operatorname{Re}(T_{\rm out})+{\rm i}\operatorname{Im}(T_{\rm out})$. \textbf{e} and \textbf{f,} The amplitude ($A(\mathbf{r})$) of the lock-in output. \textbf{g} and \textbf{h,} The phase ($\phi(\mathbf{r})$) of the lock-in output can be obtained by calculating the arctangent of $\operatorname{Im}(T_{\rm out})/\operatorname{Re}(T_{\rm out})$. \textbf{i} and \textbf{j,} The real part ($A(\mathbf{r}){\rm cos}(\mathbf{q}_{\rm CDW}\cdot\mathbf{r}+\phi(\mathbf{r}))$) of the lock-in output.}\label{EFig.5}  
\end{figure*}

\begin{figure*}
\includegraphics[width=14cm]{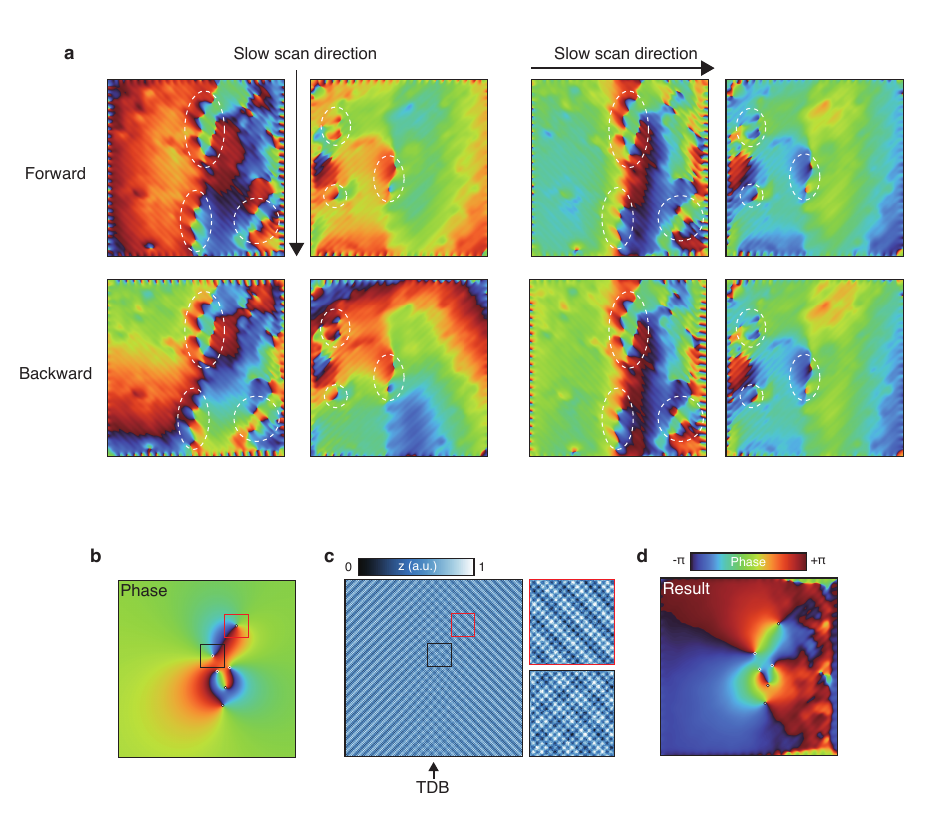}
\caption{\textbf{Proof of validity of the spatial lock-in technique.} \textbf{a,} Phase maps for the unidirectional CDWs along the $c-$(left panel) and $a-$(right panel)axes, obtained by the spatial lock-in technique. The set of phase maps in the top left corner is shown in Figures. 3(a) and 3(b). The other sets are acquired by applying the lock-in technique to the STM images for the same field of view with different scan directions (denoted in \textbf{a}). Note that all of them show topological defects at the same locations, which are marked by white dashed ellipses. To demonstrate the validity of our analysis, we construct artificial images, including two different unidirectional CDWs, TDB, and topological defects   \textbf{b,} The phase information implemented in the artificial images. We assume that the phase windings (topological defects) of the unidirectional CDW in the left domain only appear on the right side of the image. \textbf{c,} The artifical STM image is constructed with atomic corrugations, two different unidirectional CDWs, and their phase information. Each CDW is gradually attenuated across the TDB marked by a black arrow. The insets are zoomed in images for the regions marked by colored boxes, including topological defects. \textbf{d,} The phase map is acquired by analyzing the artificial STM image shown in \textbf{c} with the unidirectional CDW in the left domain as a reference signal. Note that the acquired phase map clearly identifies the topological defects that can be obscured in the STM image as shown in \textbf{c}.}\label{EFig.6}
\end{figure*}

\begin{figure*}
\includegraphics[width=14cm]{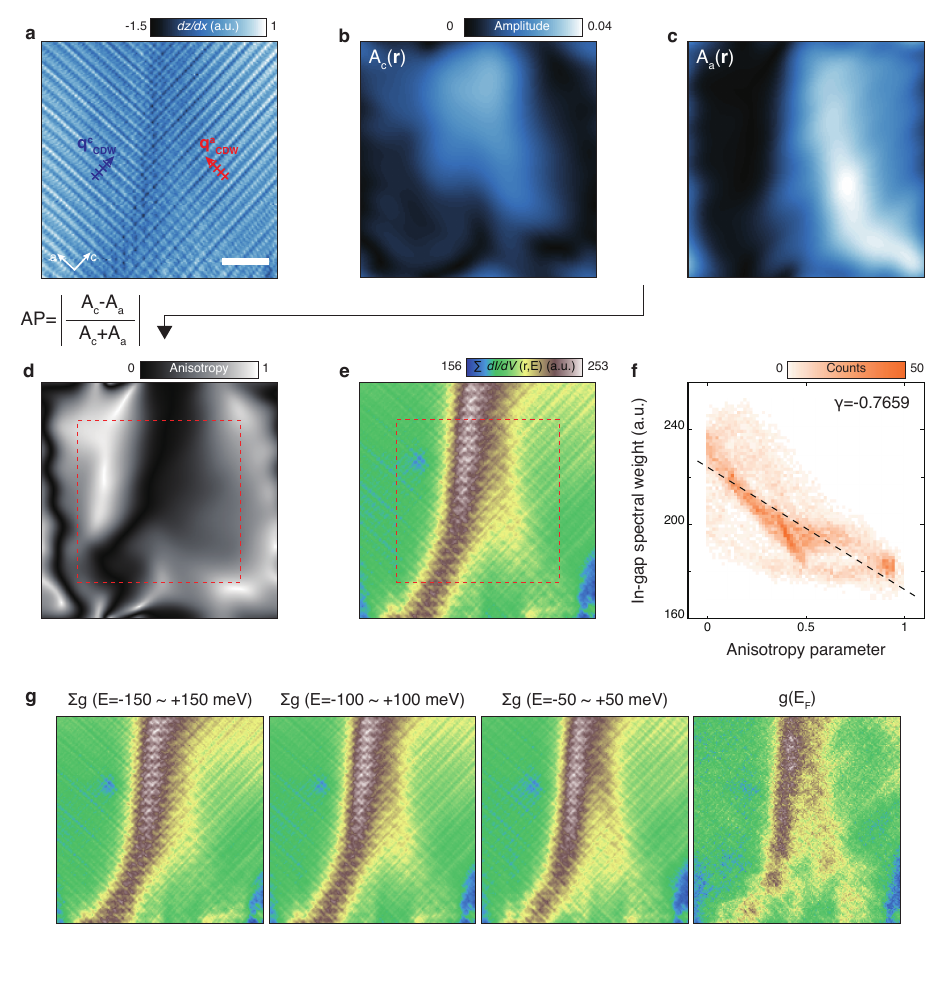}
\caption{\textbf{Determination of the anisotropy parameter.} \textbf{a,} The first derivative of the STM topographic image of the TDB shown in Figure 4(a). \textbf{b} and \textbf{c,} The amplitude ($A_{\rm c(a)}(\mathbf{r})$) maps for each unidirectional CDW. \textbf{d,} The anisotropy parameter determined by calculating the normalized difference, $\rm{AP} = \left\vert\frac{A_{\rm c} - A_{\rm a}}{A_{\rm c} + A_{\rm a}}\right\vert$, of two amplitude maps shown in \textbf{b} and \textbf{c}. \textbf{e,} In-gap spectral weight map shown in Figure 4(d). \textbf{f,} Correlation between the anisotropy parameter and the in-gap spectral weight yielding the correlation factor of -0.77. \textbf{g,} We note that the energy ranges for the determination of the in-gap spectral weight do not result in a significant difference.}\label{EFig.7}
\end{figure*}

\end{document}